%% file: AnonymousSubmission2027.tex
\documentclass[letterpaper]{article} 
\usepackage[preprint]{aaai2027}  
\usepackage[hyphens]{url}  
\usepackage{graphicx} 
\usepackage{natbib}  
\usepackage{caption} 
\usepackage{algorithm}
\usepackage{algorithmic}
\usepackage{amsmath}
\usepackage{amssymb}
\usepackage{multirow}
\usepackage{makecell}
\usepackage{array}
\usepackage{colortbl}
\usepackage{pifont}
\usepackage{comment}
\definecolor{genericblue}{HTML}{97B1CD}
\definecolor{rubricred}{HTML}{D94B55}

\newcommand{\benchyes}{\ding{51}}
\newcommand{\benchno}{\ding{55}}

\newcommand{\drop}[1]{{\scriptsize~$\downarrow$#1}}
\newcommand{\dropnote}[1]{\normalsize\textbf{$\downarrow$#1}}

\newcommand{\rqanswer}[2]{%
  \par\smallskip
  \noindent\textbf{Answer to RQ#1.} #2%
  \par\smallskip
}

\newcommand{\benchmark}{\textsc{IChart2Code}}
\newcommand{\method}{TRAIL}

\usepackage{newfloat}
\usepackage{listings}
\DeclareCaptionStyle{ruled}{labelfont=normalfont,labelsep=colon,strut=off} 
\floatstyle{ruled}
\newfloat{listing}{tb}{lst}{}
\floatname{listing}{Listing}

\usepackage{booktabs}

\title{\benchmark{}: Benchmarking Multimodal Large Language Models for Interactive Chart Code Generation}

\author{
    Xu Zhang\textsuperscript{\rm 1}\equalcontrib,
    Hongzhang Zheng\textsuperscript{\rm 1}\equalcontrib,
    Zhili Huang\textsuperscript{\rm 1},
    Yaoyi Wang\textsuperscript{\rm 1},
    Ling Xu\textsuperscript{\rm 1}\corresponding,
    Sheng Huang\textsuperscript{\rm 1}
}

\affiliations{
    \textsuperscript{\rm 1}School of Big Data and Software Engineering, Chongqing University, Chongqing, China\\
     zhangx@stu.cqu.edu.cn,
    hongzhangzheng@stu.cqu.edu.cn,
    huangzhili@stu.cqu.edu.cn,\\
    wangyy@stu.cqu.edu.cn,
    xuling@cqu.edu.cn,
    huangsheng@cqu.edu.cn
}

\begin{document}

\maketitle

\begin{abstract}
Interactive chart code generation requires models to reproduce a reference chart's appearance and underlying data and correctly implement the state changes triggered by specified user interactions. Existing chart-to-code benchmarks focus on static outputs and lack task representations or evaluation protocols for interaction specification, browser execution, and post-interaction verification. We introduce \benchmark{}, a benchmark comprising 377 tasks across 20 chart forms and 13 data families, with 
1209 interaction requirements in six families. Each task provides a reference screenshot, task-local data, and natural-language interaction requirements, with executable HTML/JavaScript code as the target output. 
We further develop a browser-based evaluation protocol with an Executability gate and three rubric-guided dimensions: Data Fidelity, Static Visual Correctness, and Interaction Correctness. The protocol tests runtime viability, consistency with task-local data, fidelity of the initial rendering to the reference screenshot, and interaction-induced state changes in a sandboxed browser. A rubric-guided MLLM judge evaluates task-specific items for the three scored dimensions using the collected browser observations and achieves an overall item-level F1 score of 0.8844 against adjudicated human labels.
 We also propose \method{}, a trajectory-guided dual-agent framework for interactive chart code generation. 
 An Inspector derives task-specific inspection checks, executes them in the browser, and uses the resulting trajectories to diagnose failures and produce structured repair feedback.
 Averaged across four MLLMs, \method{} improves the four evaluation dimensions over direct prompting by 7.89, 4.98, 3.45, and 4.47 percentage points, respectively.

\end{abstract}

\input{sections/intro}

\input{sections/relatedwork}

\input{sections/benchmark}
\input{sections/method}
\input{sections/expsetup}
\input{sections/expresult}
\input{sections/conclusion}

\bibliography{aaai2027}


\end{document}

%% file: sections/intro.tex
\section{Introduction}

\begin{figure}[t]
\centering
\includegraphics[width=\columnwidth]{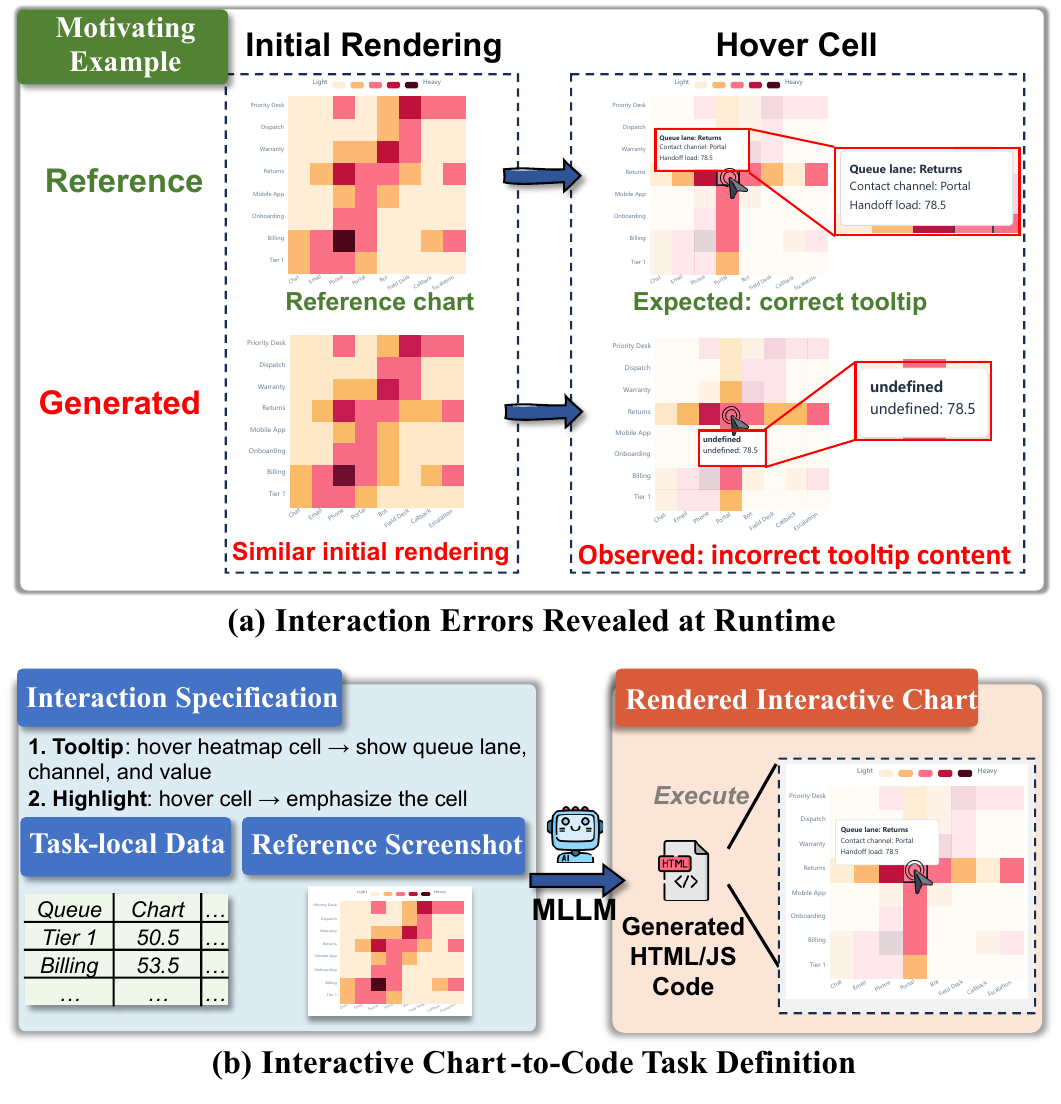}
\caption{
(a) Although the generated chart has an initial appearance similar to the reference, hovering over the same heatmap cell produces incorrect tooltip content, illustrating an interaction error that cannot be detected from the initial rendering alone.
(b) Interactive chart-to-code task definition. 
}
\label{fig:intro}
\end{figure}

\begin{table*}[t]
\centering
\small
\setlength{\tabcolsep}{3.2pt}

\begin{tabular*}{\textwidth}{
@{\extracolsep{\fill}}
l c c c c c c c c
@{}
}
\toprule

\multirow{2}{*}{\textbf{Benchmark}}
& \multicolumn{3}{c}{\textbf{Input}}
& \multirow{2}{*}{\textbf{Task}}
& \multirow{2}{*}{\textbf{Language}}
& \multirow{2}{*}{\textbf{Interactive}}
& \multicolumn{2}{c}{\textbf{Evaluation}}
\\

\cmidrule(lr){2-4}
\cmidrule(lr){8-9}

& \textbf{Img.}
& \textbf{Instr.}
& \textbf{Data}
&
&
&
& \textbf{Evaluator}
& \textbf{Rubric}
\\

\midrule

Plot2Code~\cite{wu2025plot2code}
& \benchyes
& \benchyes
& \benchno
& Reprod.
& Python/R
& \benchno
& Hybrid
& \benchno
\\

ChartMimic~\cite{shi2024chartmimic}
& \benchyes
& \benchyes
& \benchno
& Reprod.
& Python
& \benchno
& Hybrid
& \benchno
\\

ChartEdit~\cite{zhao2025chartedit}
& \benchyes
& \benchyes
& \benchno
& Edit.
& Python
& \benchno
& Hybrid
& \benchno
\\

Chart2Code~\cite{tang2025chart2code}
& \benchyes
& \benchyes
& \benchyes
& Mixed
& Python
& \benchno
& Hybrid
& \benchno
\\

RealChart2Code~\cite{zhang2026realchart2code}
& \benchyes
& \benchyes
& \benchyes
& Mixed
& Python
& \benchno
& Hybrid
& \benchno
\\

ChartEditBench~\cite{kapadnis2026charteditbench}
& \benchyes
& \benchyes
& \benchno
& Edit.
& Python
& \benchno
& Hybrid
& \benchyes
\\

\midrule

\textbf{\benchmark{}}
& \benchyes
& \benchyes
& \benchyes
& Reprod.
& HTML/JS
& \benchyes
& Hybrid
& \benchyes
\\

\bottomrule
\end{tabular*}
\caption{
Representative chart-to-code benchmarks.
Img./Instr./Data: reference image, natural-language instruction, and separately provided task-local data;
Task: Reprod. (reproduction), Edit. (editing), or Mixed (both);
Interactive: runtime end-user interaction with the rendered chart;
Evaluator: MLLM-only judging or Hybrid (automatic metrics with MLLM judging);
Rubric: task-specific evaluation criteria.
}
\label{tab:benchmark_comparison}
\end{table*}

Interactive charts support data exploration through actions such as hovering, clicking, selecting, filtering, and toggling legends, which can reveal additional values, highlight subsets, or update the displayed state. Hovering and clicking are among the most common interactions in web-based visualizations~\cite{bako2022streamlining,heer2012interactive,satyanarayan2017vegalite}. Generating an interactive chart thus requires the code to reproduce the initial view, faithfully represent the underlying data, and implement the specified responses to user actions. As shown in Figure~\ref{fig:intro}(a), the generated heatmap closely matches the reference in its initial state, yet hovering over the same cell yields incorrect tooltip content. Because this error is invisible in the initial rendering, evaluation must examine both the initial state and the behavior produced by the specified interactions.

Recent multimodal large language models (MLLMs) have demonstrated strong capabilities in visual understanding and visually grounded code generation~\cite{wang2024qwen2vl,li2024llavaonevision,yun2024web2code,si2025design2code}. Existing chart-to-code benchmarks primarily focus on static chart generation~\cite{shi2024chartmimic,zhao2025chartedit,tang2025chart2code,zhang2026realchart2code}, typically requiring models to generate code whose initial rendering matches a reference chart. Their evaluation mainly considers code executability and the structural or visual similarity of the initial rendering. Beyond chart-specific benchmarks, recent work on web code generation has begun to evaluate functional or dynamic behavior through executable tests and temporally observed outputs~\cite{lu2025webgenbench,zhang2025artifactsbench}. Evaluating interactive chart code generation further requires representing each interaction as an action, a target, and an expected outcome, executing the corresponding action in a browser, and verifying the resulting state change. However, existing chart-to-code benchmarks do not provide a chart-specific task representation and execution-based protocol that explicitly associates each interaction with its target and expected post-interaction behavior.

To address this gap, we introduce \textbf{\benchmark{}}, a benchmark for interactive chart code generation comprising 377 tasks across 20 chart forms and 13 data families, with 1209 interaction requirements organized into six families. Each task provides an interaction specification, a reference screenshot, and task-local data, and requires models to generate executable HTML/JavaScript code that reproduces the target chart and implements the specified interactions.
Figure~\ref{fig:intro}(b) summarizes this task definition. The accompanying browser-based evaluation protocol first applies an Executability gate and then scores three rubric-guided dimensions: Data Fidelity, Static Visual Correctness, and Interaction Correctness.
For each generated implementation, the evaluator runs the code in a sandboxed browser, applies the Executability tests, executes task-specific interactions, and collects static and runtime evaluation evidence. 
A rubric-guided MLLM judge then assigns binary judgments to the applicable rubric items using the collected evidence, while auxiliary automatic metrics provide complementary diagnostics. 
Against adjudicated human annotations, the rubric-guided judge shows close agreement at both the item and task levels.


Figure~\ref{fig:intro}(a) shows that interaction failures may remain hidden in the initial rendering and emerge only along the state-transition trajectory induced by a user action. This observation motivates the use of interaction trajectories as structured signals for failure localization and iterative refinement. 
We therefore propose \textbf{\method{}} (\textbf{T}rajectory-guided \textbf{R}efinement through \textbf{A}gentic \textbf{I}nspection \textbf{L}oops), a dual-agent framework for interactive chart code generation. In each bounded inspection--refinement loop, a Developer generates or revises the chart code, while an Inspector derives task-specific static and interaction checks, invokes browser tools to execute the specified chart interactions, and diagnoses the candidate from the resulting interaction trajectories.
The Inspector then provides structured repair feedback to guide the Developer's revision.
Across four MLLMs, \method{} improves most evaluation dimensions over direct prompting and yields more reliable gains than budget-matched self-refinement.

In summary, our main contributions are as follows:
\begin{itemize}
\item
We introduce \benchmark{}, a benchmark comprising 377 interactive chart code-generation tasks across 20 chart forms and 13 data families, together with a browser-based evaluation protocol that combines an Executability gate with three rubric-guided dimensions.

\item
We propose \method{}, a trajectory-guided dual-agent framework in which a Developer generates and refines interactive chart code, while an Inspector analyzes browser-executed interaction trajectories and provides structured  repair feedback for refinement.

\item
Through experiments across four MLLMs, we evaluate the effectiveness of \method{}, quantify the agreement between rubric-guided evaluation and human annotations, and characterize common failure modes in interactive chart code generation.
\end{itemize}

%% file: sections/relatedwork.tex
\section{Related Work}

\paragraph{Chart-to-Code Benchmarks.}
Existing chart-to-code benchmarks cover visual reconstruction, instruction-guided editing, and mixed generation settings. Plot2Code and ChartMimic focus on reconstructing charts from visual references; ChartEdit and ChartEditBench focus on chart editing; and Chart2Code and RealChart2Code cover mixed settings with task-local data~\cite{wu2025plot2code,shi2024chartmimic,zhao2025chartedit,kapadnis2026charteditbench,tang2025chart2code,zhang2026realchart2code}.
Table~\ref{tab:benchmark_comparison} summarizes these primarily static benchmarks.
Related work also studies interactive webpage generation, dynamic chart understanding, and dashboard reconstruction through active target exploration~\cite{xiao2025interaction2code,huang2026chartact,niu2026dashboard2code}.
In contrast, \benchmark{} evaluates generated implementations against explicit interaction requirements.

\paragraph{MLLM-as-a-Judge Evaluation.}
MLLM-based evaluators provide semantic judgments for open-ended multimodal outputs that conventional metrics cannot fully capture~\cite{chen2024mllmjudge,lee2024prometheus,xiong2025llava,ku2024viescore}. 
Recent studies further show that dynamic interaction evidence can reveal functional behaviors that static observations alone cannot capture~\cite{yang2023intercode,li2025webdevjudge,lei2026webcompassmultimodalwebcoding, meng2026webriserequirementinducedstateevaluation, wu2026benchmarkingmultimodalllmscode}.
Accordingly, our protocol grounds task-specific, rubric-guided MLLM judgments in browser-executed interaction evidence and complements them with auxiliary automatic metrics.

%% file: sections/benchmark.tex
\section{The \benchmark{} Benchmark}

\begin{figure*}[t]
    \centering
    \includegraphics[width=\textwidth]{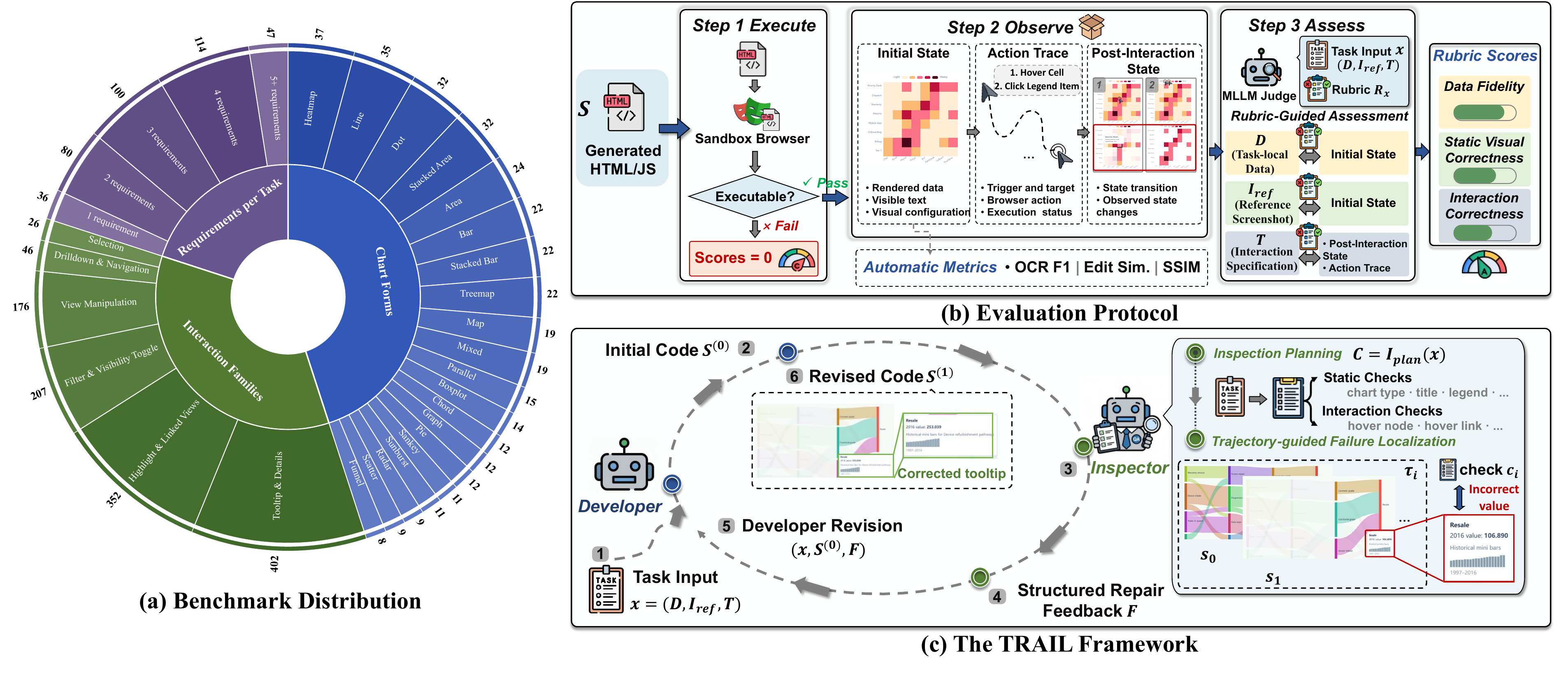}
    \caption{ 
    Overview of \benchmark{} and the \method{} framework.
    (a) Benchmark distributions over chart forms, interaction families, and requirements per task.
    (b) The browser-based evaluation protocol for \benchmark{}.
    (c) The Developer--Inspector refinement loop in \method{}.
    }
    \label{fig:framework}
\end{figure*}

\subsection{Task Definition}
Given a multimodal input $x=(I_{\mathrm{ref}},T,D)$, where $I_{\mathrm{ref}}$ is a reference screenshot, $T$ is a natural-language interaction specification, and $D$ is the set of task-local data files, an MLLM $M$ generates chart code $S$, as illustrated in Figure~\ref{fig:intro}(b):
\[
S = M(x) = M(I_{\mathrm{ref}},T,D).
\]
The output $S$ must be a single HTML file with embedded JavaScript and CSS that can be executed in a browser. A valid implementation must faithfully represent the data in $D$, reproduce the reference chart in its initial state, and implement the interactions specified in $T$.

\subsection{Benchmark Construction}
We first curate publicly accessible interactive charts with recoverable data and observable interactions, normalize them into structured seed specifications, and generate diverse task variants while removing near-duplicates. 
For each task, an MLLM produces an initial single-file HTML implementation, which human annotators execute and revise until it faithfully represents the task-local data, matches the intended initial design, and implements the specified interactions.

Three annotators then construct a task-specific rubric tree
$\mathcal{R}_x$ for the three scored dimensions: Data Fidelity, Static Visual Correctness, and Interaction Correctness.
Each dimension contains task-specific binary rubric items. Items under Data Fidelity and Static Visual Correctness specify an evaluation aspect and its expected condition, whereas each Interaction Correctness item specifies a trigger, a target, an effect, and an expected outcome.

Each task is additionally paired with four binary Executability tests that verify offline loading, the absence of critical runtime errors, visible rendering, and a nonblank chart. 
A task is retained only if its reference implementation passes all Executability tests and rubric items. Its reference screenshot $I_{\mathrm{ref}}$, interaction specification $T$, and task-local data $D$ are then packaged as $x=(I_{\mathrm{ref}},T,D)$.




\subsection{Data Statistics and Diversity}

\begin{table}[t]
\centering
\small
\setlength{\tabcolsep}{4pt}
\renewcommand{\arraystretch}{1.08}

\begin{tabular*}{\columnwidth}{@{\extracolsep{\fill}}lr@{}}
\toprule
\textbf{Statistic} & \textbf{Value} \\
\midrule

\multicolumn{2}{c}{\textbf{Input Statistics}} \\
\midrule
Tasks & 377 \\
Chart forms & 20 \\
Data families & 13 \\
Task-local data files & 419 (avg. 1.11) \\

\midrule
\multicolumn{2}{c}{\textbf{Interaction Statistics}} \\
\midrule
Interaction families & 6 \\
Interaction requirements & 1209 (avg. 3.21) \\
Tasks with multiple interaction families & 334 \\

\midrule
\multicolumn{2}{c}{\textbf{Evaluation Statistics}} \\
\midrule
Executability tests & 1,508 (avg. 4.00) \\
Data Fidelity items & 2,215 (avg. 5.88) \\
Static Visual Correctness items & 4,626 (avg. 12.27) \\
Interaction Correctness items & 3,887 (avg. 10.31) \\
Total tests and rubric items & 12,236 (avg. 32.46) \\
\bottomrule
\end{tabular*}

\caption{
Statistics of \benchmark.
Avg. denotes the average number per task.
}
\label{benchmark_statistics}
\end{table}

As summarized in Table~\ref{benchmark_statistics}, \benchmark{} contains 377 tasks spanning 20 chart forms, 13 data families, and six interaction families. 
Most tasks combine multiple interaction families,
and each task is paired with task-local data assets, Executability tests, and task-specific rubric items. Overall, the benchmark provides 10,728 rubric items across the three scored dimensions and 1,508 Executability tests.

\subsection{Evaluation Protocol}
Given a generated implementation $S$, the task input $x=(I_{\mathrm{ref}},T,D)$, and its task-specific rubric tree $\mathcal{R}_x$, the evaluator first applies the Executability tests as a hard gate and then scores three rubric-guided dimensions: \textbf{Data Fidelity}, \textbf{Static Visual Correctness}, and \textbf{Interaction Correctness}. 
The gate requires $S$ to load offline without critical runtime errors and render a visible, nonblank chart. Data Fidelity assesses consistency with $D$, Static Visual Correctness compares the initial rendering with $I_{\mathrm{ref}}$, and Interaction Correctness verifies whether each specified trigger acts on the intended target and produces the expected effect and outcome while preserving unaffected chart properties.

For each scored dimension $m \in \{\mathrm{Data},\mathrm{Static},\mathrm{Inter}\}$, let $\mathcal{R}_x^m$ denote its applicable rubric items and $y_i \in \{0,1\}$ the binary judgment for item $r_i$. 
Let $e(S)\in\{0,1\}$ indicate whether $S$ passes the Executability gate. The task-level score is
\[
\mathrm{Score}_{m}(x,S)
=
e(S)\,
\frac{1}{\lvert\mathcal{R}_x^{m}\rvert}
\sum_{r_i\in\mathcal{R}_x^{m}} y_i .
\]
Here, $e(S)=1$ if $S$ passes the Executability gate and $e(S)=0$ otherwise.

For candidates that pass the gate, Playwright~\cite{microsoft_playwright} captures the initial rendering and browser state. 
For each Interaction Correctness item, an interaction resolver maps its trigger and target to executable browser actions. The evaluator executes these actions and collects the action trace, post-interaction rendering, runtime state, and observed state changes. An evidence-grounded MLLM judge then evaluates each applicable rubric item against the task input and the collected evaluation evidence and assigns a binary judgment~\cite{chen2024mllmjudge,gu2024surveyjudge}.
We additionally report OCR F1, OCR edit similarity, and SSIM for the initial rendering as auxiliary diagnostic metrics~\cite{levenshtein1966binary,wang2004ssim}; these metrics do not replace the task-specific rubric scores.

%% file: sections/method.tex
\section{The {\method} Framework}
\subsection{Overview}

As illustrated in Figure~\ref{fig:framework}(c), \method{} consists of two agents: a Developer that generates and revises interactive chart code, and an Inspector that localizes implementation failures and provides structured repair feedback.

\subsection{Developer Agent}
The Developer is responsible for both initial code generation and feedback-guided revision. 
Given the public task input
$x=(I_{\mathrm{ref}},T,D)$, it first generates an initial implementation:
\[
S^{(0)}=\mathcal{D}_{\mathrm{gen}}(x).
\]
During refinement, the Developer receives the original implementation $S^{(0)}$ and the structured repair feedback $F$ produced by the Inspector, and generates a revised implementation:
\[
S^{(1)}
=
\mathcal{D}_{\mathrm{rep}}\!\left(x,S^{(0)},F\right).
\]

\subsection{Inspector Agent}
The Inspector follows a decoupled inspection paradigm that separates inspection planning from failure localization. 
It translates task requirements into structured static and interaction checks and uses browser-executed interaction trajectories to guide code revision.
The Inspector derives the inspection plan $C$ solely from the public task input; $C$ is distinct from the benchmark rubric $\mathcal{R}_x$, which is used only for final evaluation.
The Inspector does not access the reference implementation, evaluation-judge outputs, or private benchmark-construction metadata.

\paragraph{Inspection planning.}
Given the public task input $x$, the Inspector constructs a task-specific inspection plan
\[
\mathcal{C}
=
\mathcal{I}_{\mathrm{plan}}(x)
=
\mathcal{C}_{\mathrm{stat}}
\cup
\mathcal{C}_{\mathrm{inter}}.
\]
Each static inspection check specifies an observable property of the initial rendering and its expected condition, whereas each interaction inspection check specifies a trigger, a semantic target, an expected effect, and an expected outcome.

\paragraph{Trajectory-guided failure localization.}
To operationalize the inspection plan, we build a browser-based inspection environment that constructs a check-indexed observation bundle for each check $c_i$:
\[
\mathcal{O}_i
=
\mathcal{E}\!\left(x,S^{(0)},c_i\right).
\]
For a static check, the environment assembles $\mathcal{O}_i$ from the initial rendering and relevant task context. 
For an interaction check, it compiles the specified interaction into browser actions, executes them through the browser backend, and derives $\mathcal{O}_i$ from the resulting trajectory
\[
\tau_i=(s_0,a_1,s_1,\ldots,a_T,s_T),
\]
where $s_0$ is the pre-interaction state, $a_t$ denotes an executed browser action, and $s_t$ denotes the observable chart state after that action. 
The resulting inspection observation bundle captures the renderings and state transitions associated with $c_i$.

The Inspector combines these check-indexed observations with a bounded implementation context $\Gamma(S^{(0)})$, comprising an HTML excerpt and
compact summaries of document structure, data access, and interaction handlers, to localize failures and generate structured repair feedback:
\[
F
=
\mathcal{I}_{\mathrm{loc}}
\left(
x,
\Gamma(S^{(0)}),
\{(c_i,\mathcal{O}_i)\}_{i=1}^{N}
\right).
\]
The interaction trajectories reveal discrepancies between expected and observed behavior, while the implementation context helps identify their likely causes in the generated program. 
The resulting structured repair feedback associates each diagnosed failure with its corresponding inspection check and provides a targeted repair instruction for the Developer.

%% file: sections/expsetup.tex
\section{Experimental Setup}
\subsection{Compared Methods}
We compare \method{} against four general code-generation baselines, covering single-pass prompting and iterative refinement strategies.

\begin{itemize}
\item \textbf{Direct.} The model generates the final HTML code in a single pass from the task instruction, reference image, and public task data. 

\item \textbf{CoT}~\cite{wei2022chain}. The model is prompted to reason about the chart structure, data mapping, and interaction requirements before generating the final HTML code. 

\item \textbf{SCoT$^{*}$}~\cite{li2023scot}. We adapt structured chain-of-thought prompting to interactive chart code generation. Before producing the final HTML, the model sequentially considers the data schema, visual structure, interaction behavior, implementation structure, and consistency constraints. 

\item \textbf{Self-Refine$^{*}$}~\cite{madaan2023selfrefine}. We retain the original generation, self-critique, and revision procedure and perform one self-critique and revision cycle. The critique criteria cover executability, data fidelity, static visual fidelity, and interaction behavior.

\end{itemize}

\subsection{Experimental Configuration}
We evaluate four MLLMs: three proprietary models, including Gemini 3.1 Flash-Lite, GPT-5.4 mini~\cite{openai2026gpt54mini}, and Qwen3-VL-Plus, and the open-weight Qwen3-VL-8B-Instruct model~\cite{bai2025qwen3vltechnicalreport}. The proprietary models are accessed through their APIs, whereas Qwen3-VL-8B-Instruct is deployed locally. We use Qwen3.6-Flash as the MLLM judge for rubric-guided evaluation.

Unless otherwise specified, the code-generation models use a temperature of 0.2, a top-$p$ value of 0.95, and a maximum output length of 8,192 tokens. For \method{}, the maximum number of repair rounds is set to $K=1$, corresponding to one initial generation followed by at most one evidence-guided refinement round.



\subsection{Research Questions}
We address the following research questions:
\begin{itemize}
    \item \textbf{RQ1: Overall Performance.}
    How do different MLLMs and generation or refinement strategies perform on interactive chart code generation?

    \item \textbf{RQ2: Evaluation Validity.}
    How closely do rubric-guided MLLM judgments agree with human judgments?

    \item \textbf{RQ3: Failure Mode Analysis.}
    What failure modes occur most frequently in MLLM-generated interactive chart code?
\end{itemize}

%% file: sections/expresult.tex
\section{Experimental Results}

\begin{table*}[t]
\centering
\small
\setlength{\tabcolsep}{4pt}

\begin{tabular*}{\textwidth}{@{\extracolsep{\fill}}llccccccc}
\toprule
\multirow{2}{*}{\textbf{Model}}
& \multirow{2}{*}{\textbf{Method}}
& \multicolumn{4}{c}{\textbf{Primary Metrics (\%)}}
& \multicolumn{3}{c}{\textbf{Automatic Metrics}} \\
\cmidrule(lr){3-6} \cmidrule(lr){7-9}
& & \textbf{Exec.$\uparrow$}
& \textbf{Data$\uparrow$}
& \textbf{Static$\uparrow$}
& \textbf{Interact.$\uparrow$}
& \textbf{OCR F1$\uparrow$}
& \textbf{Edit Sim.$\uparrow$}
& \textbf{SSIM$\uparrow$} \\
\midrule
\multirow{5}{*}{\makecell{Gemini 3.1 Flash-Lite}}
& Direct & 96.02 & \underline{76.60} & 59.42 & \textbf{78.38} & 0.5429 & 0.5153 & 0.6938 \\
& CoT & \underline{96.29} & 75.84 & 60.14 & 77.27 & 0.5504 & 0.5225 & \underline{0.6974} \\
& SCoT$^{*}$ & 96.02 & 75.63 & \underline{61.50} & 76.13 & \underline{0.5588} & \underline{0.5299} & 0.6957 \\
& Self-Refine$^{*}$ & 85.94 & 66.78 & 58.40 & 72.93 & 0.5170 & 0.5010 & 0.6193 \\
& \method{} (Ours) & \textbf{96.82} & \textbf{77.38} & \textbf{61.85} & \underline{77.88} & \textbf{0.5612} & \textbf{0.5408} & \textbf{0.6987} \\
\midrule

\multirow{5}{*}{\makecell{GPT-5.4 mini}}
& Direct & 84.62 & 67.21 & 58.08 & 63.10 & 0.5130 & \underline{0.4948} & 0.6081 \\
& CoT & 81.70 & 63.28 & 52.90 & 59.51 & 0.4810 & 0.4617 & 0.5907 \\
& SCoT$^{*}$ & 82.23 & 64.84 & 54.61 & 61.74 & 0.4897 & 0.4711 & 0.5941 \\
& Self-Refine$^{*}$ & \underline{89.39} & \underline{70.77} & \underline{62.58} & \underline{68.10} & \underline{0.5150} & 0.4932 & \underline{0.6468} \\
& \method{} (Ours) & \textbf{98.94} & \textbf{77.06} & \textbf{66.45} & \textbf{74.91} & \textbf{0.5863} & \textbf{0.5690} & \textbf{0.7100} \\
\midrule

\multirow{5}{*}{\makecell{Qwen3-VL-Plus}}
& Direct & 84.62 & 60.84 & 47.63 & \underline{60.91} & 0.5043 & 0.4643 & 0.6148 \\
& CoT & 85.68 & 59.97 & 47.67 & 60.83 & 0.5006 & 0.4697 & 0.6218 \\
& SCoT$^{*}$ & \underline{86.74} & \underline{63.27} & \textbf{48.25} & 58.59 & \underline{0.5137} & \underline{0.4714} & \underline{0.6312} \\
& Self-Refine$^{*}$ & 74.01 & 52.55 & 41.00 & 47.00 & 0.4374 & 0.4030 & 0.5385 \\
& \method{} (Ours) & \textbf{89.39} & \textbf{63.92} & \underline{47.68} & \textbf{62.42} & \textbf{0.5225} & \textbf{0.4896} & \textbf{0.6477} \\
\midrule

\multirow[c]{8}{*}{\shortstack[l]{Qwen3-VL-8B-\\Instruct}}
& Direct & 58.36 & 35.43 & \underline{24.71} & 31.29
& 0.3099 & 0.2829 & 0.4276 \\
& CoT & 54.38 & 35.40 & 22.66 & 31.50
& 0.2705 & 0.2437 & 0.3967 \\
& SCoT$^{*}$ & 59.68 & \underline{39.13} & 24.32
& \underline{33.46} & 0.3140 & 0.2787 & 0.4360 \\
& Self-Refine$^{*}$ & \underline{61.27} & 37.82 & 24.44
& 31.99 & \underline{0.3181} & \underline{0.2836}
& \underline{0.4478} \\
& \method{} (Ours)
& \textbf{70.03} & \textbf{41.65} & \textbf{27.66}
& \textbf{36.35} & \textbf{0.3758} & \textbf{0.3335}
& \textbf{0.5135} \\

& \hspace{0.6em}w/o Exec. Diagnostics
& 58.36 & 37.34 & 26.29 & 32.07
& 0.3311 & 0.3085 & 0.4280 \\

& \hspace{0.6em}w/o Impl. Context
& 55.97 & 36.49 & 21.74 & 30.27
& 0.2857 & 0.2660 & 0.4054 \\

& \hspace{0.6em}w/ Unstructured Feedback
& 49.60 & 30.66 & 22.15 & 26.54
& 0.2920 & 0.2642 & 0.3612 \\

\bottomrule
\end{tabular*}

\caption{Performance comparison of prompt-based and iterative refinement
strategies across four MLLMs, with component ablations of \method{}
on Qwen3-VL-8B-Instruct.
w/o Exec. Diagnostics: removes execution diagnostics while
retaining browser-executed trajectories;
w/o Impl. Context: removes $\Gamma(S^{(0)})$ from failure
localization;
w/ Unstructured Feedback: replaces check-linked structured
repair feedback with plain-text feedback.
Boldface and underlining denote the best and
second-best results, respectively, among the five complete methods;
ablation variants are excluded from the ranking.}
\label{tab:mainRQ1}
\end{table*}


\subsection{RQ1: Overall Performance}

Table~\ref{tab:mainRQ1} compares four MLLMs under three prompt-based strategies and two iterative refinement strategies using the Executability pass rate, three rubric-guided scores, and three automatic metrics. 
All methods are evaluated on the same 377 tasks, and the two iterative refinement methods use the same maximum number of refinement rounds.

\textbf{Results.}
\method{} achieves the best result in 14 of the 16 primary model--metric comparisons and in all 12 automatic-metric comparisons.
Averaged across the four MLLMs, it improves Executability, Data Fidelity, Static Visual Correctness, and Interaction Correctness over Direct by 7.89, 4.98, 3.45, and 4.47 percentage points, respectively.
However, Static Visual Correctness remains below 70\% in every configuration, indicating that fine-grained visual reconstruction remains a major bottleneck.

The gains vary across models and metrics, with larger improvements for GPT-5.4 mini and Qwen3-VL-8B-Instruct and two metric-specific exceptions. 
Under the same one-round refinement budget, \method{} outperforms Self-Refine in all 28 comparisons, suggesting that execution-grounded diagnosis provides a more effective refinement signal than self-critique in the evaluated setting.

\textbf{Ablation Study.}
All three ablations degrade all seven metrics on Qwen3-VL-8B-Instruct. 
Removing execution diagnostics returns Executability to the Direct level of 58.36\%, while removing implementation context reduces Static Visual Correctness and Interaction Correctness by 5.92 and 6.08 percentage points, respectively. 
Replacing check-linked structured repair feedback with unstructured feedback yields the weakest ablation result on five metrics. 
These results indicate that the three components are complementary, with structured repair feedback affecting the broadest range of metrics.

\begin{figure*}[t]
    \centering
    \includegraphics[
        width=\textwidth
    ]{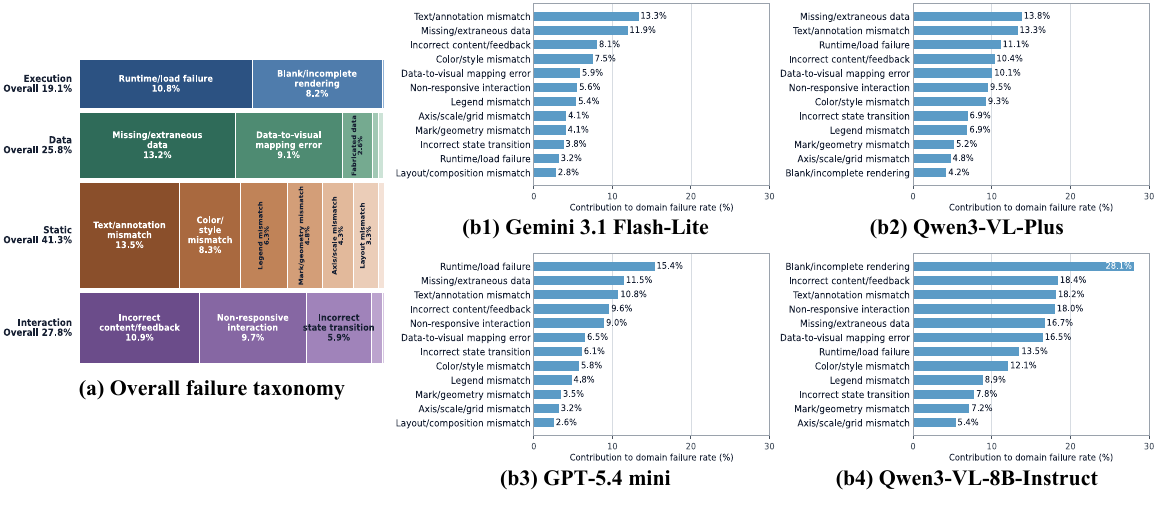}
    \caption{
    Failure analysis under the Direct setting.
    (a) Overall taxonomy: row height denotes the domain failure rate, while block area denotes a subcategory's additive contribution.
    (b1)--(b4) Leading failure modes by model; bars show task-normalized contributions to the corresponding domain failure rates.
    }
    \label{fig:failure-analysis}
\end{figure*}

\subsection{RQ2: Evaluation Validity}

\begin{table}[t]
\centering
\small
\begin{tabular*}{\columnwidth}{@{\extracolsep{\fill}}lccc}
\toprule
\textbf{Category} & \textbf{Precision} & \textbf{Recall} & \textbf{F1} \\
\midrule
Data Fidelity             & 0.7747 & 0.9180 & 0.8403 \\
Static Visual Correctness & 0.9442 & 0.9195 & 0.9317 \\
Interaction Correctness   & 0.8332 & 0.9027 & 0.8665 \\
Overall                   & 0.8579 & 0.9126 & 0.8844 \\
\bottomrule
\end{tabular*}
\caption{Item-level agreement between the rubric-guided MLLM judge and adjudicated human labels.}
\label{tab:judge_agreement}
\end{table}

\begin{table}[t]
\centering
\small
\setlength{\tabcolsep}{5pt}
\begin{tabular*}{\columnwidth}{@{\extracolsep{\fill}}lccc}
\toprule
\textbf{Category} & \textbf{Pearson} $\uparrow$ & \textbf{CCC} $\uparrow$ & \textbf{MAE} $\downarrow$ \\
\midrule

\multicolumn{4}{c}{\textbf{Generic Judge}} \\
\midrule
Data Fidelity             & 0.7936 & 0.7868 & 16.84 \\
Static Visual Correctness & 0.8211 & 0.8159 & 15.35 \\
Interaction Correctness   & 0.7407 & 0.7011 & 19.80 \\

\midrule
\multicolumn{4}{c}{\textbf{Rubric-guided Judge}} \\
\midrule
Data Fidelity             & 0.9461 & 0.9452 & 6.64 \\
Static Visual Correctness & 0.9589 & 0.9578 & 3.61 \\
Interaction Correctness   & 0.9220 & 0.9167 & 9.20 \\

\bottomrule
\end{tabular*}
\caption{Task-level score agreement with human evaluation under generic and rubric-guided judging.}
\label{tab:judge_validity}
\end{table}

\subsubsection{Item-Level Agreement with Human Annotations.}

We sample 100 tasks under the Direct setting and collect outputs from Gemini 3.1 Flash-Lite and Qwen3-VL-8B-Instruct for each task, yielding 200 generated charts and 5,790 evaluation instances from 2,895 unique rubric items across the three scored dimensions.
Three human annotators independently assigned \textsc{pass}/\textsc{fail} labels to each rubric-item instance using the same task-specific rubrics as the MLLM judge. 
Disagreements were resolved through adjudication to obtain the final reference labels. Because Executability is determined directly from runtime execution rather than by the MLLM judge, this analysis focuses on Data Fidelity, Static Visual Correctness, and Interaction Correctness.

Table~\ref{tab:judge_agreement} reports item-level agreement between the rubric-guided MLLM judge and the adjudicated human labels, with \textsc{pass} treated as the positive class. The Overall row reports micro-averaged results across all rubric items.
The judge achieves an overall precision of 0.8579, recall of 0.9126, and F1 score of 0.8844, showing close agreement with the human reference labels. Static Visual Correctness achieves the highest F1 score of 0.9317, followed by Interaction Correctness at 0.8665 and Data Fidelity at 0.8403. Data Fidelity achieves high recall (0.9180), while its comparatively lower precision (0.7747) suggests that fine-grained fidelity assessment remains more challenging.



\subsubsection{Rubric-Guided versus Generic Judging.}
We compare rubric-guided itemized judging with generic holistic judging on the same 200 human-annotated task--output instances. Both settings use the same underlying MLLM and receive identical task inputs and evaluation observations; only the rubric-guided setting receives task-specific rubric items. The generic judge directly assigns a score from 0 to 100 for Data Fidelity, Static Visual Correctness, and Interaction Correctness. 
For each task and category, the rubric-guided and human scores are computed as the percentage of corresponding rubric items labeled \textsc{pass}, yielding scores on the same 0--100 scale.

For each task and category, we compare the judge score with the corresponding human score using Pearson correlation, Lin's concordance correlation coefficient (CCC), and mean absolute error (MAE). Pearson measures linear association, CCC additionally captures deviation from the identity line, and MAE measures absolute score error; higher Pearson and CCC and lower MAE indicate closer agreement.

\textbf{Results.} Table~\ref{tab:judge_validity} reports the task-level results. Across the three categories, rubric-guided judging improves Pearson correlation by 0.1378--0.1813 and CCC by 0.1419--0.2156, while reducing MAE by 10.20--11.74 points. Static Visual Correctness achieves the closest agreement with human scores, with a Pearson correlation of 0.9589, a CCC of 0.9578, and an MAE of 3.61. Interaction Correctness remains the least aligned category, but rubric guidance still increases its Pearson correlation from 0.7407 to 0.9220 and reduces its MAE from 19.80 to 9.20. Overall, task-specific rubric itemization yields consistently closer score-level agreement with human evaluation than generic holistic judging.

\subsection{RQ3: Failure Mode Analysis}
We analyze the 1,508 HTML candidates generated by the four models under the Direct setting to characterize pre-refinement failures in a common generation setting. Executability is assessed on all 1,508 candidates, whereas the other three categories are evaluated conditionally on the 1,220 candidates that pass the Executability gate.
For each rubric-scored category, an executable candidate is counted as failing if at least one applicable rubric item receives a \textsc{fail} judgment, equivalently, if
$\mathrm{Score}_{m}(x,S)<1$. Failed rubric items are deterministically mapped to fine-grained failure modes using their item, aspect, and effect metadata. Each candidate that fails the Executability gate is assigned exactly one mutually exclusive primary cause.

\textbf{Results.} Figure~\ref{fig:failure-analysis} summarizes the overall category-level failure rates and the model-specific contributions of fine-grained failure modes. The Executability failure rate is 19.1\%, computed over all 1,508 candidates. Among the 1,220 executable candidates, the failure rates are 25.8\% for Data Fidelity, 41.3\% for Static Visual Correctness, and 27.8\% for Interaction Correctness. Static Visual Correctness therefore has the highest failure rate among the three scored categories. Within each category, the leading fine-grained modes are runtime/load failure for Executability, missing/extraneous data for Data Fidelity, text/annotation mismatch for Static Visual Correctness, and incorrect interaction content/feedback for Interaction Correctness.

The dominant failure mode also differs across models. Text/annotation mismatch occurs most frequently for Gemini 3.1 Flash-Lite, whereas runtime/load failure is the leading mode for GPT-5.4 mini. Missing/extraneous data is the most frequent mode for Qwen3-VL-Plus, while blank/incomplete rendering occurs most frequently for Qwen3-VL-8B-Instruct.



%% file: sections/conclusion.tex
\section{Conclusion}
Initial-state rendering alone is insufficient for evaluating interactive chart code because interaction-specific failures may emerge only after execution. 
\benchmark{} operationalizes this observation through task-specific interaction specifications, browser execution, and rubric-guided assessment. 
Its validation results show that itemized, evidence-grounded judging aligns more closely with human evaluation than generic holistic scoring.  
\method{} further leverages browser-executed trajectories to detect interaction failures, localize their causes, and guide targeted code revision.
The results suggest that interaction trajectories can provide actionable feedback for failure localization and iterative refinement in interactive chart code generation.

Future work will extend \benchmark{} beyond interactive chart reproduction to more complex scenarios, including instruction-guided chart editing, multi-step chart reasoning, and compositional interaction workflows.
We will also investigate how runtime interaction trajectories can support evaluation and refinement across these broader tasks.